\documentclass[conference,10pt]{IEEEtran}
\usepackage{algorithm}
\usepackage[noend]{algpseudocode}

\usepackage{lscape}

\usepackage[english]{babel}
\usepackage[T1]{fontenc}
\usepackage{cite,url,color} 
\usepackage{graphics,amsfonts}
\usepackage{epstopdf}
\usepackage{caption}
\usepackage[pdftex]{graphicx}
\usepackage[cmex10]{amsmath}
\usepackage[utf8]{inputenc}
\usepackage{enumitem}

\usepackage[font=footnotesize]{caption}
\usepackage[font=footnotesize]{subcaption}

\usepackage{tikz}
\usetikzlibrary{automata,positioning,chains,shapes,arrows,patterns}
\usepackage{pgfplots}
\usetikzlibrary{plotmarks}
\newlength\fheight
\newlength\fwidth
\pgfplotsset{compat=newest}
\pgfplotsset{plot coordinates/math parser=false}

\usepackage{array}

\usepackage[top=1.5cm, bottom=2cm, right=1.6cm,left=1.6cm]{geometry}
\usepackage{indentfirst}

\usepackage{times}

\usepackage{breqn}
\usepackage{booktabs}
\usepackage{etoolbox}
\usepackage{placeins}
\IEEEoverridecommandlockouts

\begin{document}

\tikzstyle{startstop} = [rectangle, rounded corners, minimum width=2cm, minimum height=0.5cm,text centered, draw=black]
\tikzstyle{io} = [trapezium, trapezium left angle=70, trapezium right angle=110, minimum width=3cm, minimum height=1cm, text centered, draw=black]
\tikzstyle{process} = [rectangle, minimum width=2cm, minimum height=0.5cm, text centered, draw=black, alignb=center]
\tikzstyle{decision} = [ellipse, minimum width=2cm, minimum height=1cm, text centered, draw=black]
\tikzstyle{arrow} = [thick,<->,>=stealth]
\tikzstyle{line} = [thick,>=stealth]
\tikzstyle{darrow} = [thick,<->,>=stealth,dashed]
\tikzstyle{sarrow} = [thick,->,>=stealth]
\tikzstyle{larrow} = [line width=0.1mm,dashdotted,->,>=stealth]

\title{X-TCP: A Cross Layer Approach for TCP\\Uplink Flows in mmWave Networks\vspace{-0.6ex}}

\author{\IEEEauthorblockN{Tommy Azzino, Matteo Drago, Michele Polese, Andrea Zanella, Michele Zorzi}
\IEEEauthorblockA{Department of Information Engineering, University of Padova -- Via Gradenigo, 6/b, 35131 Padova, Italy\\e-mail: \{azzinoto, dragomat, polesemi, zanella, zorzi\}@dei.unipd.it
}\thanks{The work of M. Zorzi was partially supported by NYU Wireless.}}

\makeatletter
\patchcmd{\@maketitle}
  {\addvspace{0.5\baselineskip}\egroup}
  {\addvspace{-1.3\baselineskip}\egroup}
  {}
  {}
\makeatother

\maketitle

\begin{abstract}
Millimeter wave frequencies will likely be part of the fifth generation of mobile networks and of the 3GPP New Radio (NR) standard. MmWave communication indeed provides a very large bandwidth, thus an increased cell throughput, but how to exploit these resources at the higher layers is still an open research question. A very relevant issue is the high variability of the channel, caused by the blockage from obstacles and the human body. This affects the design of congestion control mechanisms at the transport layer, and state-of-the-art TCP schemes such as TCP CUBIC present suboptimal performance. In this paper, we present a cross layer approach for uplink flows that adjusts the congestion window of TCP at the mobile equipment side using an estimation of the available data rate at the mmWave physical layer, based on the actual resource allocation and on the Signal to Interference plus Noise Ratio. We show that this approach reduces the latency, avoiding to fill the buffers in the cellular stack, and has a quicker recovery time after RTO events than several other TCP congestion control algorithms.
\end{abstract}

\begin{picture}(10,-330)(10,-340)
\put(0,0){
\put(0,0){\scriptsize The final version of this paper was accepted for presentation at} 
\put(0,-10){\scriptsize the 2017 16th Annual Mediterranean Ad Hoc Networking Workshopks.}}
\end{picture}

\section{Introduction}\label{sec:intro}
The most widely used transport protocol in the Internet is TCP, and the performance of any end-to-end connection strongly depends on how much the TCP congestion control algorithms are able to exploit the resources available at the lower layers~\cite{floyd2003highspeed}. Recently, mmWave communications emerged as a candidate to meet the 5G design goals in terms of throughput~\cite{rappaport2013millimeter}; however, the interaction with higher layer transport protocols must be carefully studied in order to avoid wasting the huge bandwidth available at the mmWave frequencies.

MmWave links are characterized by a very high throughput in Line of Sight (LOS) conditions, but the  Signal to Interference plus Noise Ratio (SINR) can decrease up to 30 dB in Non LOS (NLOS) state, because of blockage from both the human body and other obstacles (e.g., buildings, cars, people)~\cite{rappaport2}. In some cases,  a complete blockage of the signal (outage) can generate packet losses and trigger a TCP Retransmission Timeout (RTO).
This variability forces cellular network designers to introduce lower layer retransmission mechanisms, which have to rely on large buffers in order to avoid packet losses~\cite{polese2017mptcp}. However, the consequence is the emergence of the bufferbloat issue in mmWave networks~\cite{gettys2011bufferbloat,mmNet}. 

In fact, while large buffers can mask the short-term capacity variations due to the fluctuations of the mmWave channels in LOS conditions, on the other hand they can significantly delay the recognition of the sudden capacity drop due to the loss of the LOS. The sharp reduction of the link bitrate, in turn, will determine a quick increase of the queueing delay of the packets, thus increasing the probability of RTO events for the buffered packets. The TCP congestion-control mechanisms triggered by RTOs drastically reduce the transmit rate of the sender, which will take a relatively long time to fill again the huge capacity that, in the meantime, could have been recovered by the mmWave link.  

At the time of writing, the studies related to the interaction between TCP and mmWave networks~\cite{mmNet,menglei2016bufferbloat} are related to downlink connections, in which the UE is the receiver. However, because of the diffusion of cloud services and the increase of instant messaging and social network applications, the uplink traffic is increasing and the ratio between uplink and downlink is getting closer to 1~\cite{oueis2016uplink}. Therefore, in this paper we propose a novel cross layer approach for TCP uplink flows (X-TCP) with the aim of reducing the latency in NLOS scenario without reducing the communication throughput.  In particular, thanks to the control and measurement mechanisms of the cellular stack, the User Equipment (UE) is aware of the SINR of the link and of the amount of resources that the evolved Node Base (eNB) has allocated. These two pieces of information are combined in order to compute the size of the TCP congestion window.  In this way, the TCP sender does not send more packets than those that can actually be delivered to the eNB, preventing the increase of the queues size and the end-to-end packet delay. We test the effectiveness of the proposed scheme using the mmWave module of ns--3~\cite{mmWaveSim}, and compare the performance of this approach against TCP CUBIC and other congestion control protocols in terms of latency, throughput and fairness, in scenarios which are either randomly generated or designed to trigger an RTO. 

The rest of the paper is organized as follows. In Sec.~\ref{sec:sota} we will briefly review the state of the art on TCP congestion control algorithms and their performance over mmWave networks. Sec.~\ref{sec:crossapp} will describe the proposed cross layer approach. In Sec.~\ref{sec:results} we will analyze the performance of our scheme in different scenarios. Finally, in Sec.~\ref{sec:conclusion} we will draw our conclusions and suggest possible future works.

\section{State of the Art}\label{sec:sota}
Several evolutions of the original TCP congestion control algorithm have been proposed in order to address the limitations of the original TCP design (e.g., see \cite{callegari2014survey} for a recent survey of TCP versions). In particular, TCP NewReno~\cite{NewReno} does not fully exploit the available resources in high bandwidth networks, a problem that has driven the design of TCP HighSpeed~\cite{floyd2003highspeed}, TCP BIC~\cite{BIC} and TCP Illinois~\cite{Illinois}. The congestion control algorithm currently used in Linux distributions is TCP CUBIC~\cite{ha2008cubic}, which is a fairness-oriented version of TCP BIC. These variants share many mechanisms of the original TCP (i.e., \textit{slow start}, \textit{congestion avoidance}, \textit{fast retransmit} and \textit{fast recovery}), but differ in the way they react to packet losses. Both TCP NewReno and TCP Illinois are based on the Additive Increase Multiplicative Decrease (AIMD) paradigm, according to which the congestion window \texttt{cwnd} is divided by a factor $\beta>1$ for each packet loss, and then increased by summing a term $\alpha/$\texttt{cwnd} for each acknowledgment packet (ACK). The parameters $\alpha$ and $\beta$ are kept constant in TCP NewReno, and dynamically adapted to the estimated queueing delay in TCP Illinois, in order to have a steep increase of the congestion window when the congestion is less probable, and a more gentle increase otherwise. TCP BIC, instead, reacts to packet losses with a multiplicative decrease, but recovers from losses with  a more aggressive binary search of the optimal congestion window value, which can be tuned by several parameters. Finally, TCP CUBIC increases the congestion window over time, disregarding the ACK reception rate, but rather considering the absolute time since the last packet loss and adopting a cubic increase function for \texttt{cwnd}, thus mimicking the behavior of TCP Illinois. 

Despite the improvements brought about by these algorithms, the performance of TCP connections over mmWave wireless links is still suboptimal, as observed in some recent simulation studies. Paper~\cite{mmNet} presents the first performance analysis of TCP NewReno and TCP CUBIC over mmWave links. 
The main observation is related to the emergence of the bufferbloat phenomenon in NLOS conditions, which is caused by the large buffers at the Radio Link Control (RLC) layer. In fact, as briefly explained in the introduction, large buffers make it possible for the RLC to recover packet losses due to temporary short-term outages of the mmWave link in LOS conditions. Therefore, the congestion window at the sender keeps increasing in order to fill the very large capacity provided by mmWave links in LOS conditions. However, as the link experiences NLOS conditions and the datarate suddenly drops, the TCP sender does not immediately detect the capacity reduction and, hence, does not decrease the congestion window. Therefore,  the size of the RLC buffer and the queueing delay increase, eventually leading to RTO events. On the other hand, reducing the size of the RLC buffers would result in higher packet losses due to short-term variations of the link capacity also in LOS conditions, with a negative impact on the throughput. 
The importance of lower layer retransmissions for TCP on mmWave links is investigated in~\cite{polese2017mptcp}, where it is shown that retransmissions increase the throughput at the price of a higher latency. 

A first solution for downlink TCP streams over mmWave links has been proposed in~\cite{menglei2016bufferbloat}. Following the approach of~\cite{bufferbloat3G} and~\cite{mitigatingbufferbloat}, each UE uses information contained in the Downlink Control Information (DCI) to estimate the allocated bandwidth, in order to optimally tune the receive window. Then, according to the TCP congestion control RFC~\cite{allman2009tcp}, the remote TCP sender selects the minimum between the receive window and the congestion window, and if the latter is too large, then this mechanism actually imposes a limit on the number of packets that can be sent. This dynamic receive window approach is shown to decrease the latency, without affecting the throughput, when LOS to NLOS transitions occur. However, we observe that these improvements do not avoid the performance loss due to the sudden collapse of the congestion window after an RTO event and the relatively long time needed to reach again the full pipe capacity.

\section{Uplink Cross-Layer Congestion Control}
\label{sec:crossapp}

Motivated by the promising results obtained in~\cite{menglei2016bufferbloat}, in this section we describe a cross layer approach for TCP flows from the UE towards a remote destination in the Internet. The UE can use the information gathered from different layers in the cellular protocol stack to directly change the value of the TCP congestion window. As specified in~\cite{allman2009tcp}, this value is updated after the reception of each acknowlegment message. The basic algorithm is described with the pseudocode in Algorithm~\ref{alg:cl}. In particular, we set the congestion window to the optimal bandwidth delay product (BDP) that we estimate from the round trip time (RTT) of the end-to-end connection and the data rate provided by the mmWave link. 

\begin{algorithm} [!t]

	\caption{Cross layer congestion window update}
	\begin{algorithmic} 
	\small
		\State \textbf{initialization}
		\State $rtt_{\rm min} \leftarrow \infty$ 
		\State $\rm cwnd \leftarrow$ Maximum Segment Size (MSS)
		\State
		\For {every received ACK}
			\State estimate RTT $\hat{e}_{\rm rtt}$
			\State from the mmWave stack:
			\State \hspace{\algorithmicindent} estimate mmWave data rate $\hat{e}_{\rm datarate}$
			\State \hspace{\algorithmicindent} get SINR value $\Gamma$
			\If {$\hat{e}_{\rm rtt} < rtt_{\rm min}$}
				\State $rtt_{\rm min} \leftarrow \hat{e}_{\rm rtt}$
			\EndIf
			\State
			\If {$\Gamma \ge 0$ and $ \hat{e}_{\rm rtt} \le rtt_{\rm min} + \epsilon$}
			\State $ \rm cwnd$$\leftarrow \hat{e}_{\rm datarate} \, rtt_{\rm min}$
			\Else	
			\State $ \rm cwnd$$\leftarrow \lambda \, \hat{e}_{\rm datarate} \, rtt_{\rm min}$
			\EndIf
		\EndFor
	\end{algorithmic}
	\label{alg:cl}
\end{algorithm}

The RTT can be estimated after the reception of every ACK using the TCP timestamps and the method described in~\cite{RFC6298}. However, when the queueing delay in the RLC buffers starts increasing, then the RTT also increases and the BDP is artificially inflated. This worsens the congestion, since, if the congestion window blindly follows the BDP estimate, the increase of BDP due to the longer queueing delay would increment the sender rate, thus further exacerbating the congestion. Therefore, following the approach described in~\cite{mitigatingbufferbloat,bufferbloat3G}, we use the minimum value of the RTT, $rtt_{\rm min}$, as an estimate of the latency without buffering delays. Notice that the mmWave link latency (without retransmissions) has a very limited impact on $rtt_{\rm min}$ since it is smaller than 1~ms, so that the forwarding delays introduced by the core network and the public internet are dominant. Therefore, the mobility of the UE in a cell or across different cells has almost no effect on $rtt_{\rm min}$.

The core of the algorithm is the estimation of the data rate offered by the mmWave link at every congestion window update. We assume a 3GPP-like protocol stack, with a data plane composed by the PHY, MAC, RLC and PDCP layers. These will likely be part of the New Radio standard~\cite{ktRlc}. Moreover, the physical layer is assumed to be TDD-based, and the duration of each TDD slot is variable, according to the resource allocation decided by the eNB, as proposed in~\cite{dutta2017frame}. Each slot assigned to a UE is composed of a certain number of OFDM symbols, and is associated to a specific modulation and coding scheme (MCS). With this information, which is transmitted to the UE via DCIs, it is possible to compute the transport block (TB) size at the MAC layer, i.e., the number of bytes that can be actually transmitted in each resource. This computation accounts for the PHY layer header. Moreover, we scale the PHY data rate by the overhead introduced by the MAC, RLC, PDCP, IP and TCP headers. Finally, if the congestion on the network increases or the SINR $\Gamma$ is below a certain threshold (i.e., $\Gamma < 0$~dB), a scaling factor $\lambda \le 1$ is used in the increase of the congestion window, in order to decrease the aggressiveness of the protocol and account for the additional retransmissions performed by the RLC and MAC layers, or for the possible congestion in an intermediate link on the end-to-end path. The algorithm assumes that the path is congested when the estimated round trip time exceeds $rtt_{\rm min}$ by a certain threshold $\epsilon$, which is set to 10 ms following the approach in~\cite{menglei2016bufferbloat}. The parameter $\lambda$ depends on the scenario and on the configuration of the cellular network. In our implementation, we used a value $\lambda=0.85$, which was observed to be a good tradeoff between the throughput loss and the reduction of the queueing delay and, in turn, of the RTT. A context-based optimization of this parameter is left for future work.

In order to test the performance of the proposed approach, we conducted an extensive simulation campaign that will be described in Sec.~\ref{sec:results}.

\section{Performance Evaluation}
\label{sec:results}
\subsection{Simulation Setup}
\label{sec:simulator}
The performance evaluation campaign was conducted using the mmWave ns--3 module described in~\cite{mmWaveSim}. This tool allows the simulation of complex networks and scenarios with a very high level of detail. It is an end-to-end simulator, with a complete TCP/IP protocol suite and a mmWave cellular stack with the PHY, MAC, RLC, PDCP and RRC layers, as well as a core network model. The mmWave channel model is based on the statistical channel model presented in~\cite{rappaport2}. Moreover, thanks to the Buildings module of ns--3, it is possible to deploy obstacles and buildings randomly, and then deterministically state whether a link is in LOS or NLOS condition, so that this information is consistent as the UE moves in the scenario. The physical layer is OFDM-based, which is one candidate for NR waveforms~\cite{vihriala2015waveforms}. In particular, the highest MCS that can be selected yields a PHY data rate of 3.2 Gbit/s for the whole cell~\cite{mmWaveSim}. The scheduler used at the eNB is Round Robin.

We implemented the algorithm described in Sec.~\ref{sec:crossapp} as a TCP congestion control module.
In all the simulations, we consider uplink traffic from the UE to a remote host, connected to the core network gateway with a high-capacity wired link. The traffic model is full buffer~\cite{38913}, i.e., it always fills the transmission capacity of the TCP pipe with packets of Maximum Segment Size (MSS) equal to $L_{\rm pck}$ bytes, but we add the possibility of limiting the maximum application data rate to $R_{\rm app}$.

The main parameters of the simulations are summarized in Table~\ref{params}.

\begin{table}[t]
\setlength{\belowcaptionskip}{-0.5cm}

  \centering
  \begin{tabular}{@{}ll@{}}
  	\toprule
    Parameter & Value \\
    \midrule
	mmWave TX power & $30$ dBm \\
    mmWave carrier frequency & 28 GHz \\
    mmWave bandwidth & 1 GHz \\
	Number of subframes in one frame & 10 \\
	Length of one subframe in $\mu$s & 100 \\
	Number of OFDM symbols per slot & 24 \\
	Length of one OFDM symbol in $\mu$s & 4.16 \\
	Number of sub-bands & 72 \\

    UE speed $v$ & [1.75, 5] m/s \\
    $R_{\rm app, max}$ & [1, 2] Gbit/s \\
    $L_{\rm pck}$ & 1400 byte \\
    $\lambda$ & 0.85 \\
    $\epsilon$ & 10 ms \\
    RLC AM buffer size & 10 MB \\
    Core network latency & 1 ms \\
    Remote host latency & 10 ms \\
  \bottomrule
  \end{tabular}
  \caption{Simulation parameters}
  \label{params}
\end{table}

\begin{figure}[t!]
\setlength{\belowcaptionskip}{-0.5cm}

\centering
\begin{tikzpicture}[font=\sffamily, scale=0.5, every node/.style={scale=0.5}]
  \centering

    \node[anchor=south west,inner sep=0] (image) at (0,0) {\includegraphics[width=0.7\textwidth]{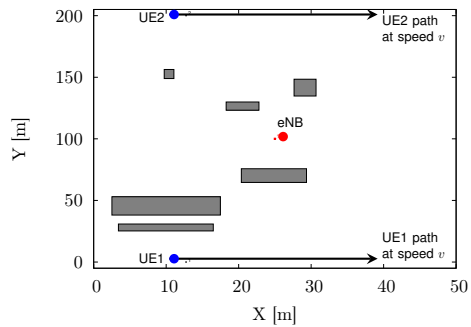}};
    \begin{scope}[x={(image.south east)},y={(image.north west)}]
        \filldraw[red,ultra thick] (0.575,0.56) circle (2pt);
        \node[anchor=south] at (0.59,0.575) (mm1label) {eNB};
        \draw[sarrow] (0.34, 0.20) -- (0.77, 0.20);
        \node[anchor=south west, text width=1.8cm] at (0.77, 0.18) (arrowLabel) {UE1 path at speed $v$};

        \filldraw[blue,ultra thick] (0.35, 0.20) circle (2pt);
        \node[anchor=east] at (0.34, 0.20) (ltelabel) {UE1};

		\draw[sarrow] (0.34, 0.92) -- (0.77, 0.92);
        \node[anchor=north west, text width=1.8cm] at (0.77, 0.93) (arrowLabel) {UE2 path at speed $v$};

        \filldraw[blue,ultra thick] (0.35, 0.92) circle (2pt);
        \node[anchor=east] at (0.34, 0.91) (ltelabel) {UE2};
    \end{scope}   
\end{tikzpicture}
\caption{First simulation scenario. The grey rectangles are randomly deployed non-overlapping obstacles (e.g., cars, buildings, people, trees).}
\label{fig:map}
\end{figure}

\begin{figure*}[t]
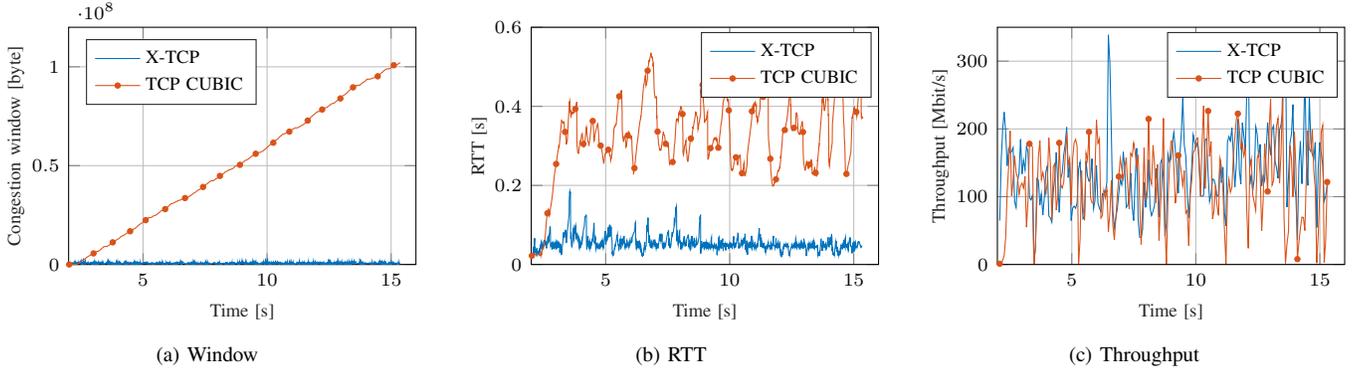

\setlength{\belowcaptionskip}{-0.2cm}

\begin{subfigure}[t]{0.33\textwidth}
\setlength{\belowcaptionskip}{-0.1cm}
	\centering
	\setlength{\fwidth}{0.8\columnwidth}
	\setlength{\fheight}{0.52\columnwidth}
	\input{Figure/window46.tex}	
	\caption{Window}
	\label{fig:clWinTime}
\end{subfigure}
\begin{subfigure}[t]{0.33\textwidth}
\setlength{\belowcaptionskip}{-0.1cm}
	\centering
	\setlength{\fwidth}{0.8\columnwidth}
	\setlength{\fheight}{0.52\columnwidth}
	\input{Figure/rtt46.tex}	
	\caption{RTT}
	\label{fig:clRttTime}
\end{subfigure}
\begin{subfigure}[t]{0.33\textwidth}
\setlength{\belowcaptionskip}{-0.1cm}
	\centering
	\setlength{\fwidth}{0.8\columnwidth}
	\setlength{\fheight}{0.52\columnwidth}
	\input{Figure/th46.tex}	
	\caption{Throughput}
	\label{fig:clThTime}
\end{subfigure}

	\caption{Evolution over time of the congestion window, RTT and throughput of a X-TCP flow and of a TCP CUBIC flow for UE1, on the path shown in Fig.~\ref{fig:map} (i.e., always NLOS).}
	\label{fig:clTime}
\end{figure*}

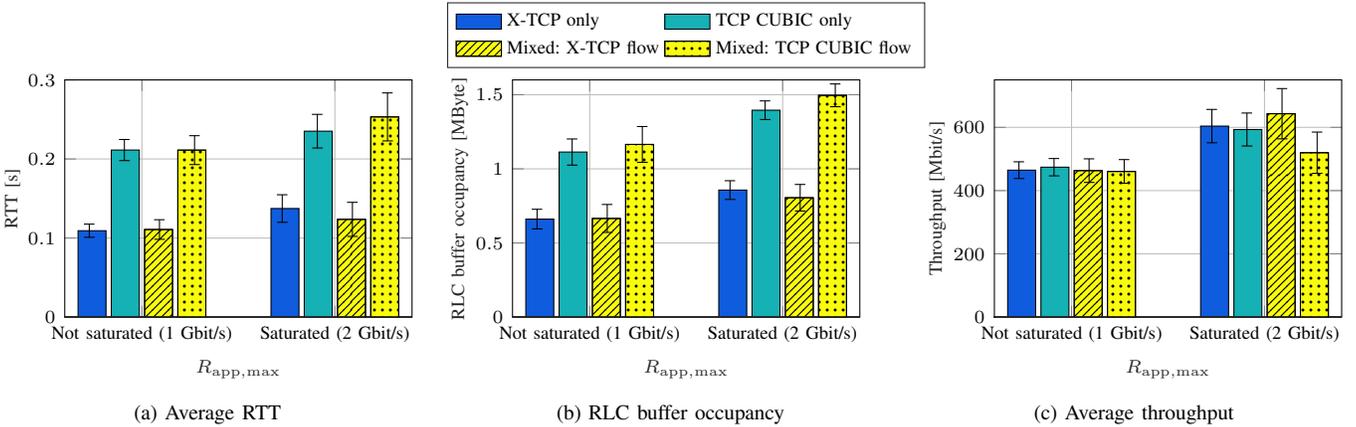
\begin{figure*}[t]
\setlength{\belowcaptionskip}{-0.4cm}

\begin{subfigure}[t]{0.33\textwidth}
\setlength{\belowcaptionskip}{-0.1cm}
	\centering
	\setlength{\fwidth}{0.8\columnwidth}
	\setlength{\fheight}{0.52\columnwidth}
%
%
\definecolor{mycolor1}{rgb}{0.05913,0.35983,0.86833}%
\definecolor{mycolor2}{rgb}{0.07227,0.48867,0.84670}%
\definecolor{mycolor3}{rgb}{0.02650,0.61370,0.81350}%
\definecolor{mycolor4}{rgb}{0.08430,0.69283,0.70617}%
\definecolor{mycolor5}{rgb}{0.34817,0.74243,0.54727}%
\definecolor{mycolor6}{rgb}{0.64730,0.74560,0.41880}%
\definecolor{mycolor7}{rgb}{0.88243,0.72743,0.32170}%
\definecolor{mycolor8}{rgb}{0.97630,0.98310,0.05380}%
\begin{tikzpicture}
\pgfplotsset{every tick label/.append style={font=\scriptsize}}

\begin{axis}[%
width=0.951\fwidth,
height=\fheight,
at={(0\fwidth,0\fheight)},
scale only axis,
bar shift auto,
xmin=0.6,
xmax=2.4,
xtick={1,2},
xticklabels={{Not saturated (1 Gbit/s)},{Saturated (2 Gbit/s)}},
xlabel style={font=\scriptsize\color{white!15!black}},
xlabel={$R_{\rm app, max}$},
ymin=0,
ymax=0.3,
ylabel style={font=\scriptsize\color{white!15!black}},
ylabel={RTT [s]},
axis background/.style={fill=white},
ylabel shift=-4pt,
xmajorgrids,
ymajorgrids,
legend style={font=\scriptsize,at={(0.05,0.85)}, anchor=south west, legend cell align=left, align=left, draw=white!15!black}
]
\addplot[ybar, bar width=0.145, fill=mycolor1, draw=black, area legend] table[row sep=crcr] {%
1	0.109276110728962\\
2	0.137343686136326\\
};
\addplot[forget plot, color=white!15!black] table[row sep=crcr] {%
0.6	0\\
2.4	0\\
};
\addlegendentry{Cross layer}

\addplot[ybar, bar width=0.145, fill=mycolor4, draw=black, area legend] table[row sep=crcr] {%
1	0.211300064942493\\
2	0.235151062691002\\
};
\addplot[forget plot, color=white!15!black] table[row sep=crcr] {%
0.6	0\\
2.4	0\\
};
\addlegendentry{CUBIC}

\addplot[ybar, bar width=0.145, fill=mycolor8, postaction={pattern=north east lines}, draw=black, area legend] table[row sep=crcr] {%
1	0.110782634090724\\
2	0.123631219313853\\
};
\addplot[forget plot, color=white!15!black] table[row sep=crcr] {%
0.6	0\\
2.4	0\\
};
\addlegendentry{Mixed, cross layer flow}

\addplot[ybar, bar width=0.145, fill=mycolor8, postaction={pattern=dots}, draw=black, area legend] table[row sep=crcr] {%
1	0.21119210101108\\
2	0.253210707194019\\
};
\addplot[forget plot, color=white!15!black] table[row sep=crcr] {%
0.6	0\\
2.4	0\\
};
\addlegendentry{Mixed, CUBIC Flow}

\addplot [color=black, draw=none, forget plot]
 plot [error bars/.cd, y dir = both, y explicit]
 table[row sep=crcr, y error plus index=2, y error minus index=3]{%
0.727272727272727	0.109276110728962	0.00833186097251123	0.00833186097251123\\
1.72727272727273	0.137343686136326	0.0174150340157094	0.0174150340157094\\
};
\addplot [color=black, draw=none, forget plot]
 plot [error bars/.cd, y dir = both, y explicit]
 table[row sep=crcr, y error plus index=2, y error minus index=3]{%
0.909090909090909	0.211300064942493	0.0133314207029255	0.0133314207029255\\
1.90909090909091	0.235151062691002	0.0212657515748233	0.0212657515748233\\
};
\addplot [color=black, draw=none, forget plot]
 plot [error bars/.cd, y dir = both, y explicit]
 table[row sep=crcr, y error plus index=2, y error minus index=3]{%
1.09090909090909	0.110782634090724	0.0123922608143934	0.0123922608143934\\
2.09090909090909	0.123631219313853	0.021546621042639	0.021546621042639\\
};
\addplot [color=black, draw=none, forget plot]
 plot [error bars/.cd, y dir = both, y explicit]
 table[row sep=crcr, y error plus index=2, y error minus index=3]{%
1.27272727272727	0.21119210101108	0.0182572449082401	0.0182572449082401\\
2.27272727272727	0.253210707194019	0.0303596412950611	0.0303596412950611\\
};
\legend{};

\end{axis}

\end{tikzpicture}%
	\caption{Average RTT}
	\label{fig:clRtt}
\end{subfigure}
\begin{subfigure}[t]{0.33\textwidth}
\setlength{\belowcaptionskip}{-0.1cm}
	\centering
	\setlength{\fwidth}{0.8\columnwidth}
	\setlength{\fheight}{0.52\columnwidth}
%
%
\definecolor{mycolor1}{rgb}{0.05913,0.35983,0.86833}%
\definecolor{mycolor2}{rgb}{0.07227,0.48867,0.84670}%
\definecolor{mycolor3}{rgb}{0.02650,0.61370,0.81350}%
\definecolor{mycolor4}{rgb}{0.08430,0.69283,0.70617}%
\definecolor{mycolor5}{rgb}{0.34817,0.74243,0.54727}%
\definecolor{mycolor6}{rgb}{0.64730,0.74560,0.41880}%
\definecolor{mycolor7}{rgb}{0.88243,0.72743,0.32170}%
\definecolor{mycolor8}{rgb}{0.97630,0.98310,0.05380}%
\begin{tikzpicture}
\pgfplotsset{every tick label/.append style={font=\scriptsize}}

\begin{axis}[%
width=0.951\fwidth,
height=\fheight,
at={(0\fwidth,0\fheight)},
scale only axis,
bar shift auto,
xmin=0.6,
xmax=2.4,
xtick={1,2},
xticklabels={{Not saturated (1 Gbit/s)},{Saturated (2 Gbit/s)}},
xlabel style={font=\scriptsize\color{white!15!black}},
xlabel={$R_{\rm app, max}$},
ymin=0,
ymax=1.6,
ylabel style={font=\scriptsize\color{white!15!black}},
ylabel={RLC buffer occupancy [MByte]},
axis background/.style={fill=white},
ylabel shift=-4pt,
xmajorgrids,
ymajorgrids,
legend style={font=\scriptsize,at={(0.5,1.05)}, anchor=south, legend cell align=left, align=left, draw=white!15!black, legend columns=2}
]
\addplot[ybar, bar width=0.145, fill=mycolor1, draw=black, area legend] table[row sep=crcr] {%
1	0.661095923167593\\
2	0.856906587774123\\
};
\addplot[forget plot, color=white!15!black] table[row sep=crcr] {%
0.6	0\\
2.4	0\\
};
\addlegendentry{X-TCP only}

\addplot[ybar, bar width=0.145, fill=mycolor4, draw=black, area legend] table[row sep=crcr] {%
1	1.11341346776236\\
2	1.39562551671719\\
};
\addplot[forget plot, color=white!15!black] table[row sep=crcr] {%
0.6	0\\
2.4	0\\
};
\addlegendentry{TCP CUBIC only}

\addplot[ybar, bar width=0.145, fill=mycolor8, postaction={pattern=north east lines}, draw=black, area legend] table[row sep=crcr] {%
1	0.665435504643624\\
2	0.805291355326136\\
};
\addplot[forget plot, color=white!15!black] table[row sep=crcr] {%
0.6	0\\
2.4	0\\
};
\addlegendentry{Mixed: X-TCP flow}

\addplot[ybar, bar width=0.145, fill=mycolor8, postaction={pattern=dots}, draw=black, area legend] table[row sep=crcr] {%
1	1.16481952931806\\
2	1.49605084641245\\
};
\addplot[forget plot, color=white!15!black] table[row sep=crcr] {%
0.6	0\\
2.4	0\\
};
\addlegendentry{Mixed: TCP CUBIC flow}

\addplot [color=black, draw=none, forget plot]
 plot [error bars/.cd, y dir = both, y explicit]
 table[row sep=crcr, y error plus index=2, y error minus index=3]{%
0.727272727272727	0.661095923167593	0.0663131893666671	0.0663131893666671\\
1.72727272727273	0.856906587774123	0.0630867300008485	0.0630867300008485\\
};
\addplot [color=black, draw=none, forget plot]
 plot [error bars/.cd, y dir = both, y explicit]
 table[row sep=crcr, y error plus index=2, y error minus index=3]{%
0.909090909090909	1.11341346776236	0.0882908222614713	0.0882908222614713\\
1.90909090909091	1.39562551671719	0.0627872190201866	0.0627872190201866\\
};
\addplot [color=black, draw=none, forget plot]
 plot [error bars/.cd, y dir = both, y explicit]
 table[row sep=crcr, y error plus index=2, y error minus index=3]{%
1.09090909090909	0.665435504643624	0.094356252957719	0.094356252957719\\
2.09090909090909	0.805291355326136	0.0894815303614047	0.0894815303614047\\
};
\addplot [color=black, draw=none, forget plot]
 plot [error bars/.cd, y dir = both, y explicit]
 table[row sep=crcr, y error plus index=2, y error minus index=3]{%
1.27272727272727	1.16481952931806	0.120670663042106	0.120670663042106\\
2.27272727272727	1.49605084641245	0.0767295202339365	0.0767295202339365\\
};
\end{axis}
\end{tikzpicture}%
	\caption{RLC buffer occupancy}
	\label{fig:clBuffer}
\end{subfigure}
\begin{subfigure}[t]{0.33\textwidth}
\setlength{\belowcaptionskip}{-0.1cm}
	\centering
	\setlength{\fwidth}{0.8\columnwidth}
	\setlength{\fheight}{0.52\columnwidth}
%
%
\definecolor{mycolor1}{rgb}{0.05913,0.35983,0.86833}%
\definecolor{mycolor2}{rgb}{0.07227,0.48867,0.84670}%
\definecolor{mycolor3}{rgb}{0.02650,0.61370,0.81350}%
\definecolor{mycolor4}{rgb}{0.08430,0.69283,0.70617}%
\definecolor{mycolor5}{rgb}{0.34817,0.74243,0.54727}%
\definecolor{mycolor6}{rgb}{0.64730,0.74560,0.41880}%
\definecolor{mycolor7}{rgb}{0.88243,0.72743,0.32170}%
\definecolor{mycolor8}{rgb}{0.97630,0.98310,0.05380}%
\begin{tikzpicture}
\pgfplotsset{every tick label/.append style={font=\scriptsize}}

\begin{axis}[%
width=0.951\fwidth,
height=\fheight,
at={(0\fwidth,0\fheight)},
scale only axis,
bar shift auto,
xmin=0.6,
xmax=2.4,
xtick={1,2},
xticklabels={{Not saturated (1 Gbit/s)},{Saturated (2 Gbit/s)}},
xlabel style={font=\scriptsize\color{white!15!black}},
xlabel={$R_{\rm app, max}$},
ymin=0,
ymax=750,
ylabel style={font=\scriptsize\color{white!15!black}},
ylabel={Throughput [Mbit/s]},
axis background/.style={fill=white},
ylabel shift=-4pt,
xmajorgrids,
ymajorgrids,
legend style={font=\scriptsize,at={(0.05,0.72)}, anchor=south west, legend cell align=left, align=left, draw=white!15!black}
]
\addplot[ybar, bar width=0.145, fill=mycolor1, draw=black, area legend] table[row sep=crcr] {%
1	464.660396363755\\
2	603.934553299993\\
};
\addplot[forget plot, color=white!15!black] table[row sep=crcr] {%
0.5	0\\
2.5	0\\
};
\addlegendentry{Cross layer}

\addplot[ybar, bar width=0.145, fill=mycolor4, draw=black, area legend] table[row sep=crcr] {%
1	474.031785699148\\
2	593.17130947336\\
};
\addplot[forget plot, color=white!15!black] table[row sep=crcr] {%
0.5	0\\
2.5	0\\
};
\addlegendentry{CUBIC}

\addplot[ybar, bar width=0.145, fill=mycolor8, postaction={pattern=north east lines}, draw=black, area legend] table[row sep=crcr] {%
1	463.156369037863\\
2	643.162078553391\\
};
\addplot[forget plot, color=white!15!black] table[row sep=crcr] {%
0.5	0\\
2.5	0\\
};
\addlegendentry{Mixed, cross layer flow}

\addplot[ybar, bar width=0.145, fill=mycolor8, postaction={pattern=dots}, draw=black, area legend] table[row sep=crcr] {%
1	460.505279196306\\
2	519.779043217859\\
};
\addplot[forget plot, color=white!15!black] table[row sep=crcr] {%
0.5	0\\
2.5	0\\
};
\addlegendentry{Mixed, CUBIC Flow}

\addplot [color=black, draw=none, forget plot]
 plot [error bars/.cd, y dir = both, y explicit]
 table[row sep=crcr, y error plus index=2, y error minus index=3]{%
0.727272727272727	464.660396363755	26.4347579773554	26.4347579773554\\
1.72727272727273	603.934553299993	52.7711342691147	52.7711342691147\\
};
\addplot [color=black, draw=none, forget plot]
 plot [error bars/.cd, y dir = both, y explicit]
 table[row sep=crcr, y error plus index=2, y error minus index=3]{%
0.909090909090909	474.031785699148	27.4033418092138	27.4033418092138\\
1.90909090909091	593.17130947336	52.1316162768954	52.1316162768954\\
};
\addplot [color=black, draw=none, forget plot]
 plot [error bars/.cd, y dir = both, y explicit]
 table[row sep=crcr, y error plus index=2, y error minus index=3]{%
1.09090909090909	463.156369037863	37.2814953529368	37.2814953529368\\
2.09090909090909	643.162078553391	79.4398869237853	79.4398869237853\\
};
\addplot [color=black, draw=none, forget plot]
 plot [error bars/.cd, y dir = both, y explicit]
 table[row sep=crcr, y error plus index=2, y error minus index=3]{%
1.27272727272727	460.505279196306	37.5539397869894	37.5539397869894\\
2.27272727272727	519.779043217859	65.2825410343807	65.2825410343807\\
};

\legend{};
\end{axis}
\end{tikzpicture}%
	\caption{Average throughput}
	\label{fig:clTh}
\end{subfigure}

	\caption{RTT, buffer occupancy and throughput for different TCP configurations. The two leftmost bars represent scenarios with two TCP flows of the same flavor, i.e., TCP CUBIC or X-TCP. The two rightmost bars report the values for a scenario with TCP flows with different flavors, i.e., one is TCP CUBIC and the other is X-TCP. The errorbars represent the 95\% confidence interval. The legend is the same for the three plots.}
	\label{fig:clThRtt}
\end{figure*}


\subsection{Evaluation in Random Scenarios}
The first scenario we consider consists of a rectangular area, with a mmWave eNB in the center and some objects (buildings, cars, people) randomly deployed over the area, without overlapping. We consider two UEs, named UE1 and UE2, moving at constant speed $v = 1.75$ m/s along straight horizontal lines. Both trajectories cross the area from left to right, but UE1 moves along the lower border of the rectangle, while UE2 is placed on the upper part of the area  (see Fig.~\ref{fig:map}). While moving, the links between the UEs and the eNB alternate between LOS and NLOS conditions, depending on the blockage due to the objects distributed in the area. 

The first set of results is reported in Fig.~\ref{fig:clTime} and shows the time evolutions of the TCP congestion window size, the RTT, and the application throughput of UE1 when using TCP CUBIC\footnote{The implementation of TCP CUBIC for ns--3 can be found at: \url{https://github.com/kronat/ns-3-dev-git/tree/tcp-versions-updated}} (marked lines) and X-TCP (plain solid lines). It can be immediately seen where the cross layer approach gains with respect to the CUBIC algorithm. As shown in Fig.~\ref{fig:clWinTime}, since the buffers and the retransmissions in the mmWave cellular stack mask most of the losses on the channel, TCP CUBIC is unaware of the actual rate provided by the lower layers and keeps increasing its congestion window, thus injecting more and more packets in the buffers at the lower layers. This yields an increase of the RTT, as shown in Fig.~\ref{fig:clRttTime}. Conversely, X-TCP keeps a small congestion window, proportional to the actual rate supported by the channel. The throughput of both approaches is instead comparable, as shown in Fig.~\ref{fig:clThTime}. In this scenario, a limited data rate $R_{\rm app, max}$ and large buffers were used. However, when the data rate is higher or the buffer size is smaller, TCP CUBIC may end up filling the whole buffer, thus causing an overflow and triggering an RTO. With X-TCP, instead, this would not happen, since the congestion window is adapted to the actual rate provided by the link.

To gain more insights on the performance of X-TCP and TCP CUBIC and on their mutual interactions, we ran a number of simulations by randomly changing the position and the number of obstacles in the area. To avoid any bias, each simulation has been run two times (i.e., resetting the pseudo-generator seed to the same value), but swapping the trajectories of the two UEs. Furthermore, we fixed the upper bound of the application-layer rate to $R_{\rm app, max} = 2 $~Gbit/s and $R_{\rm app, max} = 1 $~Gbit/s, which are larger and lower than the average bitrate of the mmWave links, respectively. 

\begin{figure*}
\setlength{\belowcaptionskip}{-0.5cm}

	\centering
	\setlength{\fwidth}{0.8\textwidth}
	\setlength{\fheight}{0.23\textwidth}
	\input{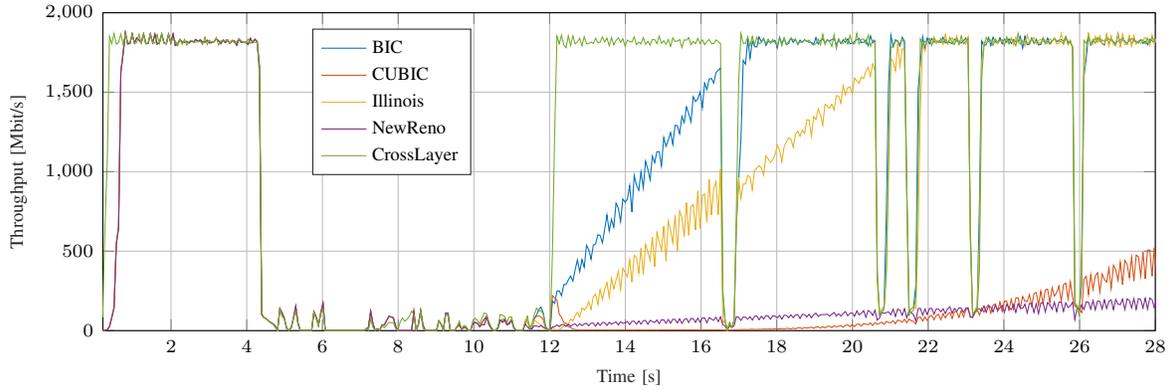}
	\caption{Throughput over time for a specific realization of the channel with the scenario in Fig.~\ref{fig:mapOutage}.}
	\label{fig:outageInTime}
\end{figure*}

Fig.~\ref{fig:clThRtt} shows the average RTT, buffer occupancy and throughput for three different combinations of TCP flavors: (i) both UEs use X-TCP, (ii) both UEs use TCP CUBIC, and (iii) one uses X-TCP and the other TCP CUBIC. For the first two cases, we show the time average of the mean performance of the two UEs, while for the latter case we present separately the results for each TCP flavor. 

\textbf{RTT and RLC buffer occupancy:} as shown in Figs.~\ref{fig:clRtt} and~\ref{fig:clBuffer}, the main advantage of the cross layer approach is the reduced latency (i.e., smaller RTT), which on average is half of that of TCP CUBIC. In particular, it can be seen in Fig.~\ref{fig:clRtt} that this behavior is consistent for the two different $R_{\rm app, max}$ values, with a slightly higher RTT for $R_{\rm app, max}= 2$ Gbit/s, and also across the three different simulated scenarios. TCP CUBIC experiences a higher RTT on average, because the congestion window continues to grow also in NLOS (as shown in Fig.~\ref{fig:clWinTime}), and packets are queued in the RLC layer buffers because the capacity offered by the mmWave link is not enough to transmit all of them. This behavior is consistent with the results in Fig.~\ref{fig:clBuffer}, where the average RLC buffer occupancy in the three scenarios is reported. The trend of this metric is similar to that of the RTT, suggesting that the RLC queueing is what causes the increase in the RTT of TCP CUBIC. X-TCP, instead, is able to adapt its congestion window to the actual rate available at the mmWave physical layer, therefore it limits the buffering at the RLC layer, only in the case of retransmissions triggered by unavoidable packet losses in the channel.

\textbf{Throughput and fairness:} it can be seen in Fig.~\ref{fig:clTh} that, when the mmWave eNB is not saturated, then there is no difference in throughput among the three scenarios. In particular, in the third scenario the aggressiveness of X-TCP does not harm TCP CUBIC, since there are enough resources available to both. When the sum of the application rates exceeds the mmWave capacity, the results in the third scenario (two concurrent flows with different TCP) suggest that X-TCP may be unfair to TCP CUBIC. When flows use the same congestion control algorithm, then the average achievable throughput is around 600 Mbit/s, independently of the TCP flavor. However, when one UE uses X-TCP and the other uses TCP CUBIC, then the two flows do not split the available resources fairly, but the cross layer congestion control algorithm achieves a 23\% higher throughput than TCP CUBIC. In particular, the TCP CUBIC flow in the third scenario loses 70 Mbit/s (i.e., 11\%) with respect to the average throughput in the second scenario (TCP CUBIC only), while the UE using X-TCP gains 50 Mbit/s (i.e., 8.5\%) with respect to the first scenario, when both UEs use X-TCP. This can be explained by the fact that, when a transition from NLOS to LOS happens, the cross layer approach restores full bandwidth utilization much more quickly than TCP CUBIC and this extra capacity is not easily released to the other flow. This has a negative impact also on the sum-rate (i.e., the sum of the throughput of the two flows), which decreases by 20 Mbit/s (i.e., 1.6\%).

\subsection{Evaluation in Blockage Scenarios}
The second simulation campaign is designed to study and compare the performance of X-TCP against that of other congestion control algorithms after an extended outage event, i.e., after a TCP retransmission timeout is triggered. These events are unlikely in a scenario with a dense deployment of mmWave base stations, because the UE would handover to a different eNB with a very high probability~\cite{poleseHo}, but they may happen nonetheless~\cite{nie201428}, thus it is important to study the behavior of transport protocols also in these scenarios. In particular, when an RTO is triggered, TCP restarts from slow start with a congestion window set to MSS bytes. In~\cite{mmNet}, a comparison between TCP NewReno and TCP CUBIC is provided. In our study, we consider both these algorithms, as well as the more aggressive TCP BIC and TCP Illinois as baselines for our comparison with X-TCP. 

\begin{figure}[t!]
\setlength{\belowcaptionskip}{-0.5cm}
\centering
\begin{tikzpicture}[font=\sffamily, scale=0.5, every node/.style={scale=0.5}]
  \centering
    \node[anchor=south west,inner sep=0] (image) at (0,0) {\includegraphics[width=0.7\textwidth]{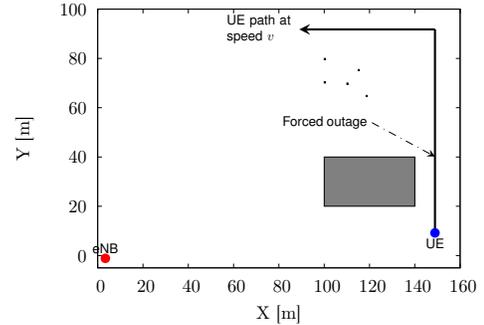}};
    \begin{scope}[x={(image.south east)},y={(image.north west)}]
        \filldraw[red,ultra thick] (0.20,0.20) circle (2pt);
        \node[anchor=south] at (0.20,0.20) (mm1label) {eNB};
        \draw[line] (0.88, 0.275) -- (0.88, 0.875);
        \draw[sarrow] (0.88, 0.875) -- (0.6, 0.875);
        \draw[larrow]  (0.75, 0.6) -- (0.88, 0.5);
		\node[anchor=east] at (0.75, 0.6) (arrowLabel2) {Forced outage};


        \filldraw[blue,ultra thick] (0.88, 0.275) circle (2pt);
        \node[anchor=north] at (0.88, 0.27) (ltelabel) {UE};

        \node[anchor=east, text width=1.8cm] at (0.6, 0.875) (arrowLabel) {UE path at speed $v$};
    \end{scope}   
\end{tikzpicture}
\caption{Outage scenario. The grey rectangles are randomly deployed non-overlapping obstacles (e.g., cars, buildings, people, trees).}
\label{fig:mapOutage}
\end{figure}

For this performance evaluation, we designed the scenario in Fig.~\ref{fig:mapOutage}, in which the UE starts in LOS, then moves to NLOS behind the gray building. After 30 meters, an outage is induced for 1 s, and the NLOS condition continues for 30 more meters. 
Finally, a LOS condition is restored. In the last part of the UE's path there are 5 small obstacles that cause temporary NLOS conditions. We consider uplink traffic with $R_{\rm app, max} = 2$ Gbit/s. The UE speed is $v = 5$ m/s.

Fig.~\ref{fig:outageInTime} shows the evolution of the throughput over time for a UE using each of the five different TCP congestion control algorithms. In the first 5 seconds, as the UE is in LOS and has a very high SINR, all the four TCP baseline versions perform the slow start phase, and then switch to congestion avoidance when the slow start threshold is reached. The congestion window continues to grow linearly, but the throughput is limited by the actual data rate available in the mmWave link, and the remaining packets are buffered at the RLC layer. X-TCP, instead, shows a steeper increase in throughput at the beginning of the simulation, since it immediately adapts to the available bandwidth, and then an approximately constant throughput, as for the baseline versions. Then, a NLOS phase begins, followed by an outage at time $t=6$ s. After the RTO expires, all the TCP congestion control algorithms restart in slow start mode, with a congestion window of one MSS, but, since the UE is still in NLOS and the SINR is low, some packets get lost, so that the congestion window does not increase and the slow start threshold is repeatedly halved. Therefore, at time $t = 12$ s, when the UE exits the NLOS condition and the available data rate increases, the slow start phase is very short, since the slow start threshold is reached immediately. The result is that TCP enters congestion avoidance too soon, and the congestion window increases very slowly, despite the large available data rate. In this phase the differences among the different congestion control algorithms are marked. X-TCP, thanks to the information on the actual rate available, almost immediately restores full bandwidth utilization. TCP BIC and TCP Illinois are the fastest in reaching high throughput values among the state-of-the-art TCP algorithms, with BIC recovering in 5~s and Illinois in 9~s. TCP CUBIC and NewReno, instead, are very slow, and do not manage to exploit the rate available before the end of the simulation. Moreover, it can be seen that the five small obstacles that the UE meets in its path do not affect the growth of the congestion window, since the lower layer buffering and retransmission mechanisms mask the conditions of the channel.

Table~\ref{table:uplinkOutage} reports the average throughput over $N=200$ simulations with the scenario of Fig.~\ref{fig:mapOutage} and an independent realization of the channel in each run. The results confirm the trend observed in Fig.~\ref{fig:outageInTime}, with TCP CUBIC and NewReno showing the lowest average throughput, because of the very slow growth of the congestion window, and X-TCP, BIC and Illinois showing the best results.

\begin{table}
\setlength{\belowcaptionskip}{-0.5cm}
\centering
\begin{tabular}{@{}ll@{}}
  	\toprule
    TCP flavor & Average throughput [Mbit/s]\\
    \midrule
	X-TCP & 1225.21 $\pm$ 15.81\\ 
	TCP BIC & 1051.32 $\pm$ 10.42\\
	TCP Illinois & 949.87 $\pm$ 10.78\\ 
	TCP CUBIC & 342.79 $\pm$ 8.46\\
	TCP NewReno & 342.46 $\pm$ 10.33\\
  \bottomrule
  \end{tabular}

	\caption{Average throughput for different congestion control algorithms for indipendent runs of the scenario in Fig.~\ref{fig:mapOutage}. The throughput is reported with the 95\% confidence interval.}
	\label{table:uplinkOutage}
\end{table}

\vspace{-0.05cm}
\section{Conclusions}\label{sec:conclusion}
In this paper we presented a cross layer approach for the dynamic adjustment of the TCP congestion window in a mmWave uplink scenario. The communication at mmWave frequencies offers a potential gigabit throughput that state of the art transport protocols cannot fully exploit because of the variable quality of the mmWave channel. In particular, two critical behaviors emerge: the bufferbloat issue and the consequent increase in the delay in NLOS conditions, and the slow recovery after extended outages that cause TCP retransmission timeouts. These are the consequences of the abstract view that TCP has of the end to end connection. In order to improve the overall performance we introduced an alternative TCP congestion control algorithm that uses the information on the SINR and the resource allocation available to the UE in order to tune the TCP congestion window to an optimal value, that minimizes the queueing delay in the RLC layer buffers without harming the throughput. The performance of this approach was tested in a randomly generated scenario, with a detailed and realistic end-to-end simulator, and we showed that the average RTT and RLC buffer occupancy are more than 50\% smaller to those of TCP CUBIC. The negative aspect of the proposed approach is a decrease in fairness with respect to legacy flows. The analysis was then extended to a scenario designed to trigger an RTO, and the performance of the cross layer congestion control algorithm was compared to that of TCP BIC, Illinois, CUBIC and NewReno, showing that the proposed approach is the fastest to reach the full utilization of the mmWave link bandwidth.

As future work, we propose to extend and refine the design of this protocol in order to account more properly for lower layer retransmissions, fully exploiting the information available in the mmWave protocol stack. In particular, we will focus on the optimization fo the parameter $\lambda$, and we will evaluate the interplay of X-TCP with other traffic models. Finally, we will introduce a TCP split approach in order to account for bottlenecks in other segments of the end-to-end connection and to make this approach feasible also for downlink scenarios.
\vspace{-0.2cm}

\bibliographystyle{IEEEtran}
\bibliography{IEEEabrv,./IEEEabrv.bib}

\end{document}